\documentclass[aps,prb,twocolumn,groupedaddress,showpacs,floatfix]{revtex4}
\usepackage{amssymb}
\usepackage{amsmath}
\usepackage{graphicx}

\newcommand{\pll}{photoluminescence }

\begin{document}

\title{Comment on: ``Luminescence spectra of quantum dots in microcavities. ''}

\author{Nicol\'as Quesada}
\email{nquesada@pegasus.udea.edu.co}
\author{Paulo C\'ardenas}
\email{pcardenas@fisica.udea.edu.co}
\author{Boris A. Rodr\'iguez}
\email{banghelo@fisica.udea.edu.co}

\address{Instituto de F\'isica, Universidad de Antioquia, AA 1226 Medell\'in,
Colombia}
\date{May 31, 2010}

\begin{abstract}
In this comment we show that there is a direct connection between coherent
exchange of energy among light and matter and the emission spectrum of a
microcavity quantum dot system as modeled in Phys. Rev. B {\bf 79}, 235325
(2009) by F. P. Laussy, E. del Valle, and C. Tejedor. To do so, we show
that in their model the necessary and sufficient conditions for
having eigenvalues with non-zero imaginary parts in the propagator of the bare
mode populations, are the same as for having strong coupling in the emission spectrum. This
amounts to saying that, whenever there is strong coupling there will be oscillating
frequencies in the dynamics of the populations. These conditions are valid both
for the case where matter is treated as bosonic or fermionic, in the
spontaneous emission case.
\end{abstract}
\maketitle


Laussy and coworkers \cite{teje2,teje3,teje1} studied the effects of
considering incoherent pumping of both excitons and photons in a microcavity
quantum dot system. In particular, they clearly show the nontrivial effects of
such pumping mechanisms in the emission spectrum of the system.  The system is modeled
using the following quantum master equation (in units where $\hbar=1$):
\begin{eqnarray}
\label{lindblad}
 \frac{d}{dt} \rho &=& i [\rho,H]+\sum_{c=a,b} \frac{\gamma_c}{2}\left(2 c 
\rho c^{\dagger}-c^{\dagger}c \rho-\rho c^{\dagger} c \right)\\
 & &+\sum_{c=a,b}
\frac{P_c}{2}\left(2 c^{\dagger} 
\rho c-cc^{\dagger} \rho-\rho c c^{\dagger}\right), \nonumber
\end{eqnarray}
where $H=\sum _{c=a,b} \omega_c c^{\dagger}c+g \left(
a^{\dagger} b+a
b^{\dagger} \right)$. $a$,$a^{\dagger}$ are photonic boson operators and $b$,
$b^{\dagger}$ can be either bosonic or fermionic matter operators. The
quantities $\omega_c$ are
the bare energies of the light and matter, $g$ is the coupling constant
between them, $\gamma_c$ are the decaying rates of the cavity and the emitter
and $P_c$ are the rates at which they are being incoherently pumped.

\begin{figure}
\includegraphics[width=0.45\textwidth]{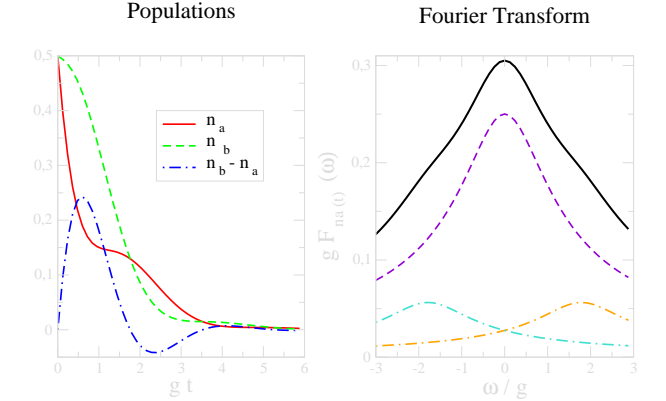}
\caption{\label{populationsplot} \label{populationsfourier}
(left panel) Time evolution of the
populations of the
photons $n_a$, excitons $n_b$  and their difference for the
parameters $\Delta=P_a=P_b=0$, $\gamma_a=1.9 g$ and
 $\gamma_b=0.1 g$. The initial condition was an upper polariton,
$\langle a^{\dagger} b \rangle (0)= n_a(0) = n_b(0)=1/2$.
(right panel) The dashed lines are the components of the Fourier transform of
the signal $n_a(t)$. The solid black line is the sum of the components
of the signal, \emph{i.e.}, the Fourier transform of $n_a(t)$.
}
\end{figure}

For the case where the exciton is modeled as a boson Laussy \emph{et. al.} define a
criteria for having strong coupling (SC)\cite{teje2}:
\begin{equation}
\label{sc}
 g>|\Gamma_-|,
\end{equation}
where
$4
\Gamma_\pm=\Gamma_a\pm\Gamma_b=(\gamma_a-P_a)\pm(\gamma_b-P_b)$.
We will show, that the same criterion of the above equation, is a necessary and
sufficient condition for having oscillatory frequencies in the propagator of the
bare mode populations. For convenience we re-write equation (12) of
reference [\onlinecite{teje2}]
as follows:
\begin{eqnarray}
\label{populations}
& & \dfrac{d}{dt} \left(
\begin{array}{c}
 n_a \\
 n_b \\
 \beta \\
 \alpha
\end{array}
\right) =
\left(
\begin{array}{c}
 P_a \\
 P_b \\
 0 \\
 0
\end{array}
\right)  + \\
& & \left(
\begin{array}{cccc}
 -\Gamma _a & 0 & -2 g & 0 \\
 0 & -\Gamma _b & 2 g & 0 \\
 g & - g & -2 \Gamma _+ & -\Delta \\
 0 & 0 &  \Delta  & -2 \Gamma _+ 
\end{array}
\right) 
\times
\left(
\begin{array}{c}
 n_a \\
 n_b \\
 \beta \\
 \alpha \nonumber
\end{array}
\right),
\end{eqnarray}
where $n_a = \langle a^\dagger a\rangle$,
$n_b = \langle b^\dagger b \rangle$,
$\alpha=\Re(\langle a^{\dagger} b \rangle)$, $\beta=\Im(\langle a^{\dagger} b \rangle)$,
and $\Delta=\omega_a-\omega_b$. 
The above equation can be written in a more compact form as: $\frac{d}{dt}
\mathbf{w}(t) =  \mathbf{f} + \mathbf{A}  \mathbf{w}(t)$, where $\mathbf{A}$
corresponds to the coefficient matrix, $\mathbf{w}(t)$ is the vector which
contains the single time mean values of interest, \emph{i.e.},
$\mathbf{w}(t)=(n_a(t),n_b(t), \beta(t),\alpha(t))^T$ and
$\mathbf{f}=(P_a,P_b,0,0)^T$. The formal solution of this equation is:
\begin{equation}
 \mathbf{w}(t)=e^{\mathbf{A}t}
\mathbf{w}(0)+\mathbf{A}^{-1}  \left( e^{\mathbf{A} t} - 1 \right) \mathbf{f} ,
\end{equation}
where $e^{\mathbf{A} t}$ is the propagator of the last equation.
The eigenvalues of the matrix $\mathbf{A}$ are given by:
\begin{eqnarray}
\lambda(\mathbf{A})_{\pm,\pm}&=&-2 \Gamma_+ \pm \sqrt{2} \sqrt{a\pm \sqrt{b}};\\
a&=&\Gamma_-^2-g^2-\frac{\Delta^2}{4}; \nonumber \\
b&=&(\Gamma_-^2-g^2)^2+(g^2+\Gamma_-^2)\frac{\Delta^2}{2} + \frac{\Delta^4}{16}. \nonumber
\end{eqnarray}
In resonance condition ($\Delta=0$) they simplify to:
\begin{equation}
 \lambda(\mathbf{A})=\{-2 \Gamma_+,-2 \Gamma_+,-2 \Gamma_++ 2 i R_0,
-2 \Gamma_+- 2 i R_0 \},
\end{equation}
where $R_0=\sqrt{g^2-\Gamma_-^2}$. Notice that if $R_0$ is a positive 
number, the condition (\ref{sc}) is automatically fulfilled, 
\emph{i.e.}, light and matter are in the SC regime. Then, if $R_0>0$, there
will be imaginary frequencies in the propagator that will lead to
oscillations in the populations. As a consequence of choosing as variables the real and imaginary parts of the
coherence $\langle a^\dagger b \rangle$, the propagator in resonance condition is given by a block diagonal matrix, with form:
\begin{eqnarray}
e^{\mathbf{A} t}=
 \left(
\begin{array}{ll}
 \mathbf{M}_{3\times 3}(t) & \mathbf{0} \\
 \mathbf{0} & e^{-2 \Gamma _+ t}
\end{array}
\right).
\end{eqnarray}
Notice that the variable $\alpha$
decouples from equation (\ref{populations}) and undergoes exponential
decay $\alpha(t)=e^{-2 \Gamma_+ t} \alpha(0)$.
The square matrix  $\mathbf{M}_{3\times 3}(t)$ is given by:

\begin{widetext}

\begin{eqnarray}
\label{propagator}
\mathbf{M}_{3\times 3}(t)&&= e^{-2 \Gamma_+ t} \times \\
&&\left(
\begin{array}{lll}
 \frac{ \left(R_0 \cos \left(R_0 t\right)-\Gamma_- \sin \left(
   R_0 t\right) \right){}^2}{R_0^2} & \frac{ g^2  
   \sin ^2\left(R_0 t\right)}{4 R_0^2} & \frac{2 g   \sin
   \left(R_0 t\right) \left( \Gamma_- \sin \left(R_0 t\right)-R_0 \cos \left(t
   R_0\right)\right)}{R_0^2} \\
 \frac{ g^2   \sin ^2\left(R_0 t\right)}{4 R_0^2} &
   \frac{ \left( \Gamma_-\sin \left(R_0 t\right)  +R_0 \cos
   \left(R_0 t\right) \right){}^2}{R_0^2} & \frac{2 g  
   \sin \left(R_0 t\right) \left( \Gamma_-\sin \left(R_0 t\right)  + R_0 \cos
   \left(R_0 t\right) \right)}{R_0^2} \\
  \frac{g  \sin \left(R_0 t\right) \left( R_0 \cos \left(
   R_0 t\right) - \Gamma_- \sin \left( R_0 t\right) \right)}{R_0^2} &
   - \frac{g   \sin \left(R_0 t\right) \left( \Gamma_-\sin
\left(
   R_0 t\right) +R_0 \cos \left(R_0 t\right) \right)}{R_0^2} &
   \frac{ \left(g^2 \cos \left(2  R_0 t\right)-\Gamma
   _-^2\right)}{R_0^2}
\end{array}
\right). \nonumber
\end{eqnarray}

\end{widetext}
The above equation explicitly shows that in SC the populations
undergo oscillations with frequencies proportional to $ 2 R_0$.\\
In this paragraph, we re-examine the results in section V of ref. 
[\onlinecite{teje2}].
In figure 6 of the same article they present the dynamics of the average photon
number $n_a$ for different initial conditions. In particular, they claim that in
strong coupling there is no oscillation in such observable for the initial
condition of one upper polariton with the parameters $\Delta=P_a=P_b=0$,
$\gamma_a=1.9 g$ and
 $\gamma_b=0.1 g$. When one plugs their
parameters into the
propagator, equation (\ref{propagator}), one sees that there are oscillating
frequencies, yet the dynamics seems to be simply decaying in the populations
$n_a$ and $n_b$. When their difference is plotted it is seen that it actually
oscillates showing that indeed both terms contain imaginary frequencies
(see the left panel of fig. \ref{populationsplot}). One might wonder why the
oscillations are not clearly visible in the populations. The reason is
that the spectrum of $\mathbf{A}$ contains two purely real
eigenvalues that will hinder the other two which are complex. This is further clarified
when one takes the Fourier transform of the propagator and applies it to a given
initial condition. In the right panel of fig. \ref{populationsfourier} we plot
the Fourier components of $n_a$ for the parameters of the left panel of the same
figure. It is clearly seen how the widened Lorentzian around zero hides the rest
of the contributions at $\pm 2 R_0 \approx  \pm 1.78606 g$. The reason why $n_b - n_a$
 clearly shows oscillations, with period $T = 2\pi /(2R_0)$, is precisely that the zero 
frequency component cancels out after the substraction.\\
Summarizing the above discussion, we have shown that for the quantum master
equation (\ref{lindblad}) the following statements are equivalent:
\begin{itemize}
 \item The propagator of the bare mode populations has imaginary frequencies.
 \item The system, as modeled by equation (\ref{lindblad}),  is in SC regime.
\end{itemize}
This implies that, whenever oscillations are observed in
the dynamics of the bare mode populations the system will be in SC regime.
In the case where one models the exciton with a fermionic operator ($b=\sigma$),
the spectrum of \pll cannot be obtained analytically when the effects of incoherent
pumping are present. When $P_a=P_b=0$ 
one can obtain exactly the eigenvalues of the
dynamical matrix involved in the equations of the quantum regression theorem.
del Valle \emph{et. al.} define $n$th order SC as the following condition \cite{teje3}:
\begin{equation}
g>|\Gamma_-|/\sqrt{n}
\end{equation}
Here, we first show that the condition above is also necessary and sufficient to
have imaginary frequencies, in the dynamics of the populations and coherences in
the same model.
To do so, we use the notation defined in [\onlinecite{perea}].
We arrange the populations and coherences of the density matrix $\rho$ as
follows:
$\mathbf{u}=(\rho_{G0,G0},\rho_{G1,G1},\rho_{X0,X0},\beta_{1},\alpha_{1},
\ldots,$ $\rho_{Gn,Gn},\rho_{Xn-1,Xn-1},\beta_n,\alpha_n,\ldots)^
T$, $\alpha_n=\Re(\rho_{Gn,Xn-1})$, $\beta_n=\Im(\rho_{Gn,Xn-1})$ and
$\rho_{im,jn}=\langle i,m|\rho|j,n\rangle$. Where $i,j$ are either $G$ (ground
state) or $X$ (exciton state) and $n,m$ are integers representing the number of
photons in the Fock state.

The equation of motion of $\mathbf{u}$ is
\begin{equation}
\label{exp}
 \frac{d}{dt} \mathbf{u} = \mathbf{B} \mathbf{u}.
\end{equation}
The structure of the matrix $\mathbf{B}$ will
consist of block diagonal terms of sizes $1\times1$ (only one block,
corresponding to the vacuum $\rho_{G0,G0}$), $4\times4$ and off
diagonal terms \emph{over} the diagonal of the matrix \cite{perea}. This implies
that
to reduce the matrix $\mathbf{B}$ to an upper triangular matrix it is necessary
to rotate each diagonal block and the blocks above it. Once the matrix has an
upper triangular form their eigenvalues are simply given by the elements of the
diagonal. Summarizing, the eigenvalues of the whole matrix $\mathbf{B}$ are
simply the eigenvalues of the blocks $1\times1$ and $4\times4$.\\
The structure of the blocks $4\times4$ corresponding to the dynamical equations
of the $n$th excitation manifold is:
\begin{widetext}
 \begin{eqnarray}
\mathbf{B}|_{4\times4}=\left(
\begin{array}{llll}
 -n \gamma_a  & 0 & -2 g \sqrt{n} & 0 \\
 0 & -((n-1) \gamma_a +\gamma_b)  & 2 g \sqrt{n} & 0 \\
  g \sqrt{n} & -g \sqrt{n} & -\frac{\gamma_b +(2 n-1) \gamma_a }{2} & -\Delta \\
 0 & 0 & \Delta & -\frac{\gamma_b +(2 n-1) \gamma_a }{2} 
\end{array}
\right).
\end{eqnarray}
\end{widetext}
The eigenvalues of the above matrix are given by:
\begin{eqnarray}
 \lambda(\mathbf{B}|_{4\times4})_{\pm,\pm}&=&-\frac{\gamma_b }{2}-n \gamma_a
+\frac{\gamma_a }{2}\pm
\sqrt{2} \sqrt{c \pm \sqrt{d}}; \nonumber \\
c&=& \Gamma_-^2 - n g^2-\frac{\Delta ^2}{4}; \nonumber \\
d&=& g^4 n^2+2 g^2 n \left(\frac{\Delta ^2}{4}- \Gamma _-^2\right) \nonumber\\
& &+\left(\frac{\Delta ^2}{4}+\Gamma _-^2\right){}^2.
\end{eqnarray}
Notice that all equations derived for fermionic matter reduce to the boson matter model when 
$n=1$ and $P_a = P_b=0$. However this correspondence breaks down with finite pumping or 
for higher excitation manifolds.
In order to have $\left. c\pm\sqrt{d}\right| _{\Delta=0}<0$
it is necessary that $g>|\gamma_a-\gamma_b|/(4\sqrt{n})=|\Gamma_-|/\sqrt{n}$.
That is, the condition for having $n$th order strong coupling is the same as
for having imaginary eigenvalues in the propagator of equation (\ref{exp}).


We would like to emphasize that the emission spectrum of the system is a robust
quantity, whereas the dynamics of the populations, in general, depends on the
initial conditions. Yet, the propagators of the dynamical systems
(\ref{populations}) and (\ref{exp}) are also robust features that show a similar
dependence on
the parameters that the emission spectrum presents.\\
 It is important to be able
to correlate the dynamical regimes (SC or weak coupling, WC) with the dynamics
of the populations and coherences. For instance, the study of the entanglement
between the exciton and photonic subsystems will require the knowledge of the
elements of the density matrix. In the SC regime one will expect entanglement
sudden death and re-birth while in WC one will see how the entanglement either
vanishes in a finite time or approaches asymptotically to zero\cite{nico}. 
Finally, we would like to emphasize that the connections found between
population dynamics and the emission spectrum of the system are particular to
the model given by equation (\ref{lindblad}). In general, one cannot expect them to
be related, and one can find models 
where the frequencies of the propagator and the first order correlation function
depend in a different way on the system parameters\cite{elena,nico1,vera}.\\

\bibliography{comm}

\end{document}